\documentclass{nseJournal}
\usepackage{local}
\usepackage{algorithm}
\usepackage{algorithmic}
\usepackage{caption}
\usepackage{hyperref}
\usepackage{ulem}
\usepackage{afterpage}
\usepackage{color}
\usepackage[numbers]{natbib}

\begin{document}

\title{A Quasi-Monte Carlo Method with Krylov Linear Solvers for Multigroup Neutron Transport Simulations
}

\addAuthor{Sam Pasmann}{a}
\addAuthor{Ilham Variansyah}{a}
\addAuthor{C. T. Kelley}{b}
\addAuthor{\correspondingAuthor{Ryan McClarren}}{a}
\correspondingEmail{rmcclarr@nd.edu}

\addAffiliation{a}{Department of Aerospace and Mechanical Engineering\\ University of Notre Dame\\ Fitzpatrick Hall, Notre Dame, IN 46556}
\addAffiliation{b}{North Carolina State University, Department of
Mathematics\\ 3234 SAS Hall, Box 8205\\ Raleigh NC 27695-8205}

\addKeyword{Neutron Transport}
\addKeyword{Monte Carlo Methods}
\addKeyword{Quasi Monte Carlo}
\addKeyword{Krylov Linear Solvers}

\titlePage

\begin{abstract}

In this work we investigate replacing standard quadrature techniques used in deterministic linear solvers with a fixed-seed Quasi-Monte Carlo calculation to obtain more accurate and efficient solutions to the neutron transport equation (NTE). Quasi-Monte Carlo (QMC) is the use of low-discrepancy sequences to sample the phase space in place of pseudo-random number generators used by traditional Monte Carlo (MC). QMC techniques decrease the variance in the stochastic transport sweep and therefore increase the accuracy of the iterative method. Historically, QMC has largely been ignored by the particle transport community because it breaks the Markovian assumption needed to model scattering in analog MC particle simulations. However, by using iterative methods the NTE can be modeled as a pure-absorption problem. This removes the need to explicitly model particle scattering and provides an  application well-suited for QMC. To obtain solutions we experimented with three separate iterative solvers: the standard Source Iteration (SI) and two linear Krylov Solvers, GMRES and BiCGSTAB. The resulting hybrid iterative-QMC (iQMC) solver was assessed on three one-dimensional slab geometry problems. In each sample problem the Krylov Solvers achieve convergence with far fewer iterations (up to 8x) than the Source Iteration. Regardless of the linear solver used, the hybrid method achieved an approximate convergence rate of $O(N^{-1})$, as compared to the expected $O(N^{-1/2})$ of traditional MC simulation, across all test problems. 
\end{abstract}

\section{Introduction}
\label{sec:intro}

Solving the neutron transport equation (NTE) under various conditions, accurately, and efficiently is vital to nuclear reaction simulations like those in advanced reactor design or accident analysis \cite{Duderstadt1977}. The neutron transport equation describes the distribution of neutrons in space, angle, energy, and time. The equation's high-dimensional nature makes it difficult to design efficient general-purpose algorithms and many solution techniques, most common of which have been stochastic Monte Carlo simulations \cite{Murray1977} or deterministic discrete ordinates ($S_N$) methods \cite{Lewis1984}, have been developed.

For deterministic solutions, Source Iteration (SI) is the simplest and most  common deterministic solution technique for solving the discrete ordinates method \cite{Lewis1984}. SI is equivalent to a fixed-point Picard Iteration, nevertheless, as problems become collision dominated, the convergence rate of the SI can become arbitrarily slow \cite{Adams2002,Warsa2004}. More advanced iteration techniques such as Krylov subspace methods, including Generalized Minimal RESidual method (GMRES) and BiConjugate Gradient STABilized method (BiCGSTAB), have been shown to outperform  standard Source Iteration, particularly when there are highly scattering materials \cite{Adams2002}. Nonetheless, as the dimensionality and fidelity of the problem increases, the deterministic quadrature techniques used to evaluate the system of equations become intractable \cite{morel2003analysis, Willert2013Thesis}. 

Monte Carlo (MC) simulations provide a more robust solution by using random sampling and probability to produce solutions. In this method the statistical error scales according to $O(N^{-1/2})$ --- where $N$ is the number of neutron histories --- regardless of the dimensionality of the problem \cite{williams2013random}. However, MC simulations are often seen as a \textit{last resort} due to their high computational cost and slow rates of convergence \cite{dupree2002monte, kalos2009monte, McClarren2018}. Recent work  by Willert et al. investigated a hybrid MC-deterministic solution where the deterministic quadrature sweep of the iterative method was replaced with a Monte Carlo transport simulation \cite{Willert2013Thesis, ctk:jeff1}. This method attempts to combine the efficiency of iterative methods while also providing an accurate solution for complex problems given the robustness of MC simulation. However, it is found that a staggering number of  particle histories are may be required for convergence of the iterative method even for mono-energetic slab problems \cite{ctk:jeff1}. Or, as Heinrich Von Kleist wrote in a previous age \cite{Kohlhaas}, ``probability is not always on the side of truth.'' 

Our work investigates the use of \textit{fixed-seed} Quasi-Monte Carlo (QMC) techniques in place of standard, pseudo-random MC to decrease the variance in the transport process and therefore improve the convergence of the iterative method. Quasi-Monte Carlo techniques use low-discrepancy sequences (LDS) in place of typical pseudo-random number generators for Monte Carlo sampling. Various LDS have been developed, including the Sobol and Halton sequences, the goal of each is to sample the phase space in a deterministic and \textit{self-avoiding} manner. Theoretically, this results in a sampling convergence rate proportional to $O(N^{-1})$, compared to $O(N^{-1/2})$ of standard Monte Carlo \cite{Bickel2009}. 

Rather than taking subsequent samples from the same LDS at the start of every iteration, the LDS is reset to the beginning of the sequence. This \textit{fixed-seed} approach allows the iterative method to converge at a much faster rate than if new samples were taken. With typical random number generators, this technique would be avoided to ensure samples are uniformly distributed through the phase-space. However, the low-discrepancy nature of the Sobol Sequence, Halton Sequence, etc. ensure a well-balanced sampling of the phase-space even with a relatively low number of samples.   

Despite the benefits offered by QMC, it has largely been ignored by the particle transport community \cite{Spanier1995}. There has been some recent work in using QMC for radiative transfer problems without scattering \cite{Farmer2020, MOROKOFF1993}, but,  to the knowledge of the authors, there has not been any recent work with QMC applied to neutron transport. This is likely because the deterministic nature of the LDS breaks the Markovian assumption needed for the particle random-walk when scattering is present.  Therefore, QMC must be implemented in applications which are not Markovian processes or steps must be take to ensure the Markovian assumption is held.

Presently, there have been two strategies for implementing QMC in particle transport. The first is known generally as randomized-QMC or (RQMC) which includes a host of strategies that attempt to randomize a sequence and still retain the low-discrepancy of the samples \cite{Fox1999, Spanier1995, MOROKOFF1993, Konzen2019}. Depending on the randomization technique, the theoretical convergence rate  of $O(N^{-1})$ may be reduced \cite{He2015convergence, Palluotto2019}. While other randomization techniques may theoretically uphold the $O(N^{-1})$ convergence rate, but are computationally expensive to execute \cite{Palluotto2019}. The second implementation of QMC in particle transport is simply in use of problems without scattering, primarily seen in radiative heat transfer problems \cite{Farmer2020, MOROKOFF1993}.

Our proposed iterative-QMC (iQMC) method allows for both: problems that include scattering and the use of unaltered LDS, i.e., no RQMC method is required. This is achieved by modeling the problem in the QMC simulation as a purely absorbing system where each particle is emitted and \textit{traced} out of the volume. After this process, or \textit{QMC Sweep} as it will be referred to from now on, the scattering term is iterated upon using a deterministic linear solver and the process repeats until a desired tolerance or maximum number of iterations is reached, thereby removing the need for the simulation of a \textit{random walk} process \cite{Pasmann2021}.

The outline of this paper is as follows: Section \ref{sec:solvers} presents a brief overview of the neutron transport equation, Source Iteration, and Krylov solvers. Section \ref{sec:QMC} describes the use of fixed-seeding and low-discrepancy sequences to form the Quasi-Monte Carlo transport sweep. Section \ref{sec:algorithm} provides an overview of the implemented algorithms before analysis and results from three 1-D test problems are presented in Section \ref{sec:results}. The first problem solves for scalar flux in an infinite medium with multi-group data generated from FUDGE \cite{mattoon2012generalized} with a known analytic solution. The second problem, known as \textit{Reed's Problem}, is a multi-media problem benchmarked with results from a high particle count Monte Carlo simulation, using the Center for Exascale Monte-Carlo for Neutron Transport's (CEMeNT's) Monte-Carlo Dynamic Code (MCDC) \cite{Variansyah2022}. The third and final problem, provided by Garcia et al., provides angular flux results at the slab edges from a fixed boundary source with a spatially decaying scattering cross section \cite{cesinh}. Finally, key findings and future work are discussed in Section \ref{sec:conclusion}.

\afterpage{\clearpage}

\section{Methods}
\label{sec:methods}

 
 \subsection{Neutron Transport Source Iteration and Krylov Methods}
 \label{sec:solvers}
 
We begin with the one-speed neutron transport equation in slab geometry with isotropic scattering \cite{Lewis1984}:
\begin{equation}\label{eq:transport}
\mu  \frac{\partial \psi}{\partial x} (x,\mu) + \Sigma_{t}(x) \psi(x,\mu) =
\frac{1}{2} \left[\Sigma_s(x) \phi(x) + q(x)\right],
\end{equation} 
\begin{equation}
	\label{eq:phi}
	\phi(x) = \int_{-1}^{1}\psi(x,\mu) d\mu,
\end{equation}
for $0 \le x \le \tau$. The boundary conditions are

\begin{equation}\label{eq:bc}
\psi(0, \mu>0) = \psi_{l}(\mu), \qquad \psi(\tau, \mu<0) = \psi_{r}(\mu).
\end{equation}
Where $x,\mu$ are the particle position and angle respectively, $\psi$ is the angular flux, $\phi$ is the scalar flux, $\Sigma_t$ is the total macroscopic cross section, $\Sigma_s$ is the scattering macroscopic cross section, and $q$ represents an internal source function.


\subsubsection{Source Iteration}

The transport Eqs. \eqref{eq:transport}--\eqref{eq:phi} can be solved iteratively via Source Iteration (SI):

\begin{equation}\label{eq:SI}
\mu  \frac{\partial \psi^{(n+1)}}{\partial x} (x,\mu) + \Sigma_{t}(x) \psi^{(n+1)}(x,\mu) =
\frac{1}{2} \left[\Sigma_{s}(x) \phi^{(n)}(x) + q(x)\right],
\end{equation}
\begin{equation}
	\phi^{(n+1)}(x) = \int_{-1}^{1}\psi^{(n+1)}(x,\mu) d\mu,
\end{equation}
where superscript $^{(n)}$ indicates iteration index. The equations can be represented in operator notation as:
\begin{equation}
    \label{eq:transport_sweep}
	\phi^{(n+1)} = \cals[\phi^{(n)}, q, \psi_l, \psi_r],
\end{equation}
where the transport sweep operator $\cals$ updates a scalar flux estimate given an internal source $q$ and boundary sources $\psi_l$ and $\psi_r$. In the proposed hybrid method, this transport sweep operation is performed via Monte Carlo simulation (which is discussed later in Sec. \ref{sec:QMC}).

The SI Equation \eqref{eq:transport_sweep} can be rewritten as follows:
\begin{equation}
    \phi^{(n+1)} = \calk[\phi^{(n)}] + f,
\end{equation}
where
\begin{equation}
    \calk[\phi^{(n)}] = \cals[\phi^{(n)}, 0, 0, 0],
\end{equation}
and
\begin{equation}
    f = \cals[0, q, \psi_l, \psi_r].
\end{equation}
By collecting the scalar flux terms, one can demonstrate that SI is equivalent to the fixed-point Picard iteration of a linear problem
\begin{equation}
    \label{eq:linear_form}
    A \phi \equiv (I - \calk) \phi = f,
\end{equation}
where $I$ is the identity function. Equation \eqref{eq:linear_form} is in a form we can send to linear solvers, particularly those that are more efficient than the fixed-point Picard iteration, such as Krylov methods. Note that we do not need to explicitly form the matrix $A$, we only need to compute the action of $A$ on $\phi$, which is accomplished with the \textit{QMC Sweep}.


 

\subsubsection{Krylov Methods}

An order-$r$ Krylov subspace is defined with notation from the previous section as \cite{ctk:roots}:
\begin{equation}
K_r = \textrm{span}(\phi, A\phi, A^{2}\phi, ..., A^{r-1}\phi).
\end{equation}

For each experiment presented in Sec. \ref{sec:results}, two Krylov methods, GMRES \cite{gmres} and BiCGSTAB \cite{bicgstab}, were used. The Generalized Minimum RESidual (GMRES) is one of the most common Krylov methods. When solving $A\vec{\phi}=\vec{f}$, GMRES minimized $||f-A\phi||_2$ over the $k^{th}$ Krylov subspace. For every iteration, the GMRES stores an additional Krylov vector. For problems that require many iterations this may lead to memory constraints. BiCGSTAB is a low-storage Krylov method that is memory bounded throughout the algorithm. However, the memory savings come from information that is thrown out with each iteration and therefore BiCGSTAB will generally require more iterations to converge than GMRES. Nonetheless, as we will observe in Sec. \ref{sec:results}, both Krylov methods will require far fewer iterations than the SI.

 
\subsection{Quasi-Monte Carlo Transport Sweep}
\label{sec:QMC}

\subsubsection{Monte Carlo Transport Sweep}

\begin{figure}[h]
\centerline{
\includegraphics[width=3.5in]{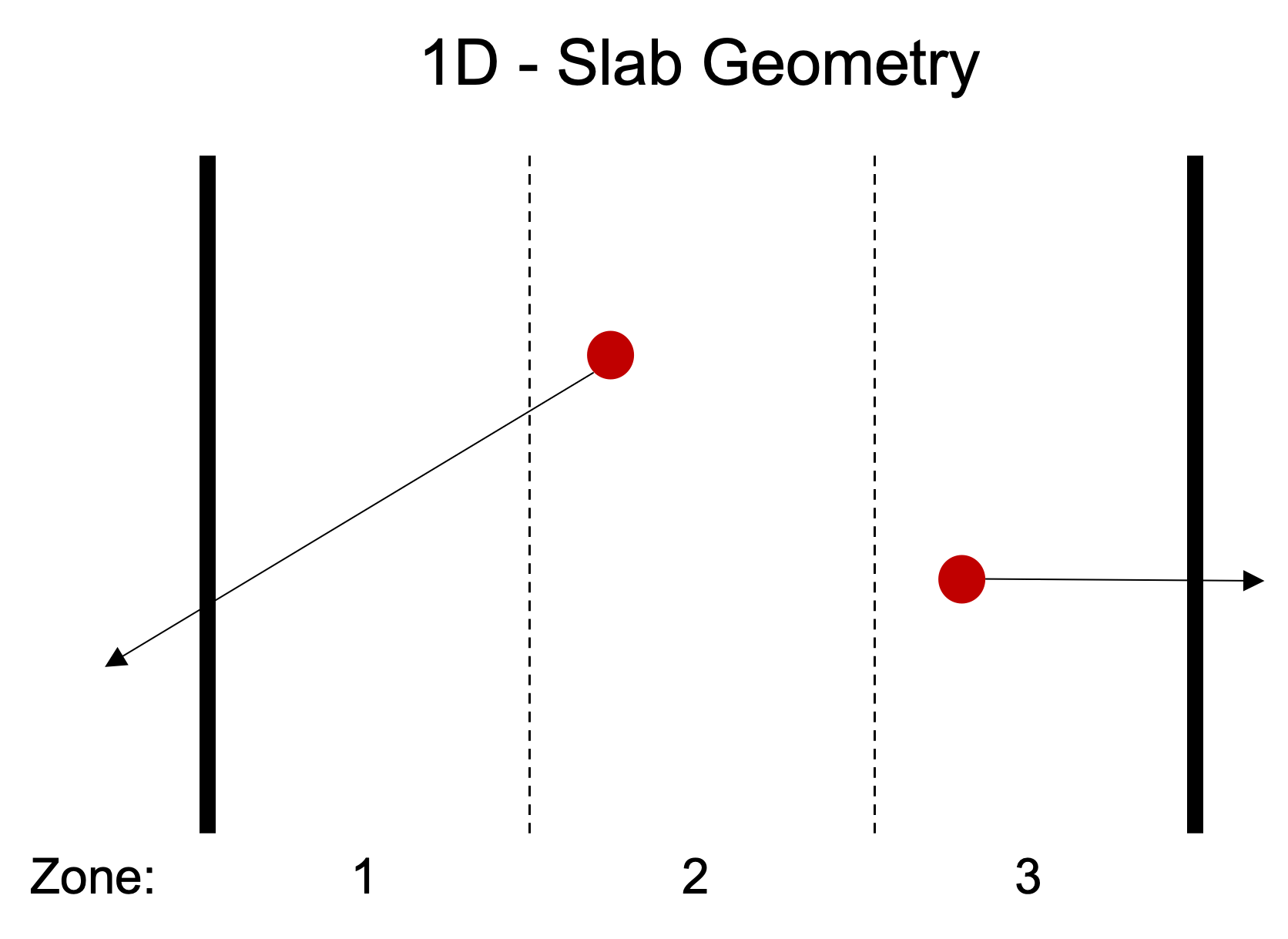}}
\caption{\label{fig:sweep} A simplified diagram of a Monte Carlo sweep. For each MC sweep the particles are emitted with initial position and angle $(x_i, \mu_i)$ and are swept out of the volume, tallying the scalar flux according to the path length tally estimator (Equation \ref{equation:tally}).}
\end{figure}

Monte Carlo methods for neutron transport seek to simulate the behavior of a statistically significant number of particles from \textit{birth} to \textit{death} to gain an approximate behavior of the system. For our one-dimensional simulations, each particle begins with an initial position ($x_i$), direction ($\mu_i$), and statistical weight ($w_i$). In an analog simulation, the particle would then be tracked from collision to collision, tallying quantities of interest as the particle moves. Each time the particle undergoes a scattering collision, a new direction ($\mu$) would be sampled and the next distance to collision would be calculated. This process would repeat until the particle is either absorbed or exits the volume. However, Eq.~\eqref{eq:SI} is a purely absorbing transport problem with a known source. MC simulation in a purely absorbing system can be enhanced by employing the continuous weight absorption technique (also called implicit capture), which continuously reduces the statistical weight of each particle per length traveled ($s$):
\begin{equation}
    \label{equation:CWR}
    w_{\textrm{new}} = w_{\textrm{old}}e^{-\Sigma_{a}s}.
\end{equation}
Consequently, after emission the particle is traced straight out of the volume reducing the statistical weight according to the distance traveled across each spatial cell, as illustrated by Figure \ref{fig:sweep}.

We use the track-length tally estimator to compute the spatially-averaged scalar flux in the defined mesh. Because the weight is continuously reduced with each step, the tally scoring becomes:
\begin{equation}
    \label{equation:tally}
    \frac{1}{V}\int_{0}^{s}w_{\textrm{old}}e^{-\Sigma_{a}s\prime}ds = \frac{w_{\textrm{old}}}{V}\left(\frac{1-e^{-\Sigma_{a}s}}{\Sigma_{a}}\right) .
\end{equation}
This estimate of the scalar flux is then used to compute the scattering source in the next iteration of the solver.


\subsubsection{Low-Discrepancy Sequence for Quasi-Monte Carlo}

Given the purely absorbing system (Equation \ref{eq:SI}) and the use of continuous weight absorption technique (Equations \ref{equation:CWR} and \ref{equation:tally}), the only things need to be randomly sampled are the particle initial position $x_i$ and direction $\mu_i$. In a standard MC transport sweep, a pseudo-random number generator is used to sample $x_i$ and $\mu_i$. In Quasi-Monte Carlo transport sweep, a quasi-random low-discrepancy sequence is used instead.

\begin{figure}[h]
\centerline{
\includegraphics[width=0.9\textwidth]{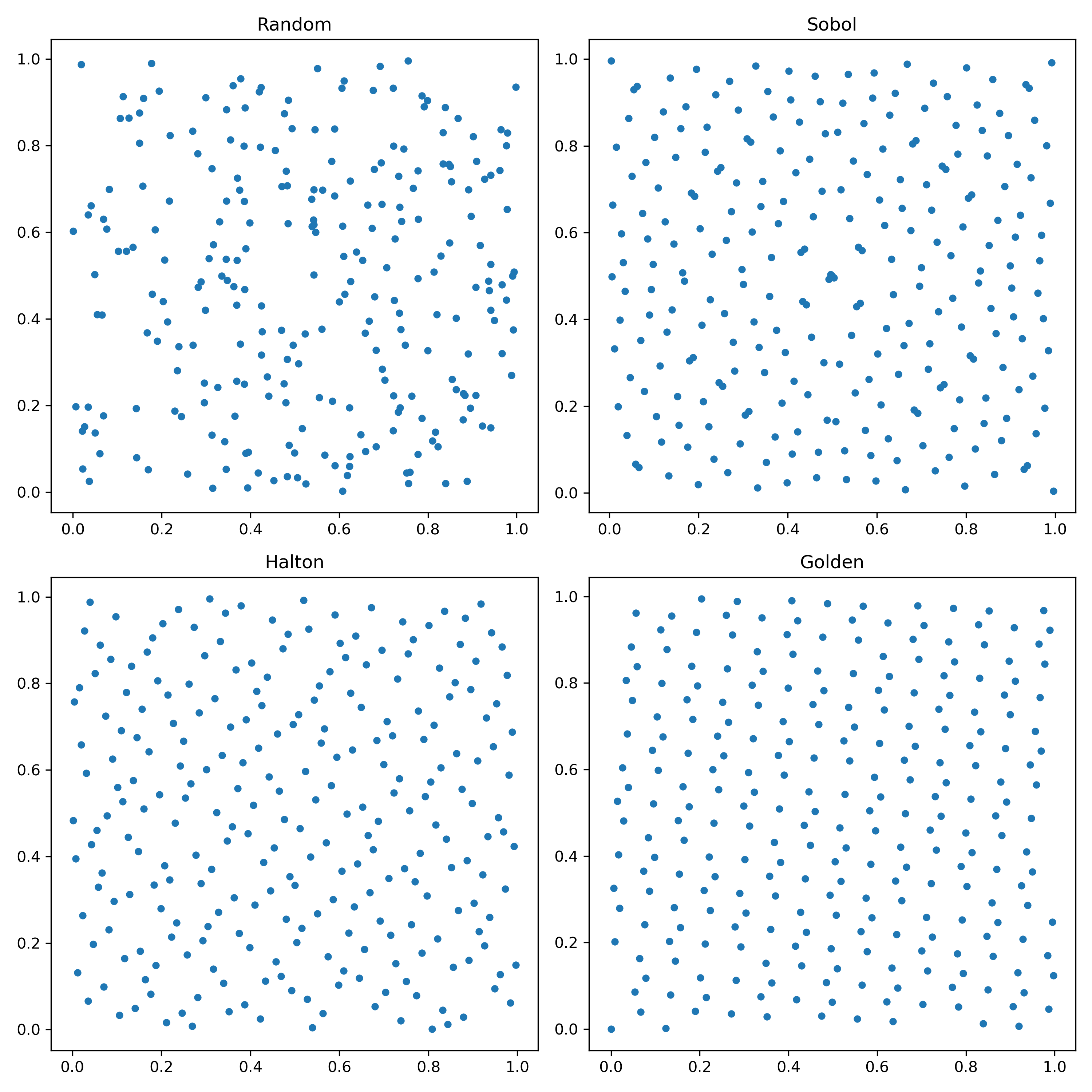}
}
\caption{\label{fig:rng_comparison} 256 points generated in a unit square with pseudo random points (top left), Sobol Sequence \cite{Johnson2020} (top right), Halton Sequence \cite{Driscoll2019} (bottom left), and Golden Sequence \cite{Moritz2021} (bottom right).}
\end{figure}

Low-discrepancy sequences use deterministic algorithms to sample the phase space in a \textit{self-avoiding} manner thereby approaching a more uniform distribution and approximating the expectation more efficiently. This results in a theoretical convergence rate of $O(N^{-1})$ compared to the $O(N^{-1/2})$ from pseudo-randomly placed points \cite{Palluotto2019}. In addition to the well-known Sobol and Halton Sequences, Figure \ref{fig:rng_comparison} shows the distribution of 256 points in a unit square of a newer LDS known as the \textit{Golden Sequence} \cite{Moritz2021}.

\subsubsection{Multigroup Vectorization}
Neutron cross sections vary greatly with energy and contain large resonance regions, making them computationally expensive to model with high fidelity. The multigroup method is a common approach used to model energy-dependent cross sections that splits the energy range into $G$ finite regions, each with a representative cross section. The multigroup equations for $G$ groups in 1D slab geometry in matrix form are \cite{Lewis1984}.

\begin{equation}
    \label{eq:MGmat}
    \mu  \frac{\partial \vec{\psi}}{\partial x} (x,\mu) + \underline{\Sigma}_{t}(x) \vec{\psi}(x,\mu) =
    \frac{1}{2}  \underline{\Sigma}_{s}(x) \int_{-1}^1 \vec{\psi}(x, \mu') \dmup + \frac{\vec{q}(x)}{2},
\end{equation}
where
\begin{equation}\label{eq:vecs}
\vec{\psi} = (\psi_1, \psi_2, \dots, \psi_G)^\mathrm{T}, \qquad \vec{q} = (q_1, q_2, \dots, q_G)^\mathrm{T}, 
\end{equation}
\begin{equation}\label{eq:MatricesT}
 \underline{\Sigma}_{t}(x)  = \begin{pmatrix} \Sigma_{t,1}(x) & 0 & \dots & 0\\
 0 & \Sigma_{t,2}(x) &  & 0 \\
 \vdots & & \ddots & \vdots\\ 
 0 & \dots & 0 & \Sigma_{t,G}(x) 
 \end{pmatrix}, 
\end{equation}
and
\begin{equation}\label{eq:MatricesS}
 \underline{\Sigma}_{s}(x)  = \begin{pmatrix} \Sigma_{s,1\rightarrow 1}(x) & \Sigma_{s,2\rightarrow 1}(x)  & \dots & \Sigma_{s,G\rightarrow 1}(x) \\
 \Sigma_{s,2\rightarrow 1}(x) & \Sigma_{s,2\rightarrow 1}(x)  & \dots & \Sigma_{s,G\rightarrow 2}(x) \\
 \vdots & \vdots & & \vdots\\
 \Sigma_{s,G\rightarrow 1}(x) & \Sigma_{s,G\rightarrow 1}(x)  & \dots & \Sigma_{s,G\rightarrow G}(x) \\
 \end{pmatrix}.
\end{equation}
As a consequence of this energy discretization and the employed scattering-free particle tracing technique, \textit{each} particle in the MC sweep can now represent \textit{all} energy groups. Conversely, in analog MC each simulated particle may only represent one energy group and random samples are taken to determine up or down scattering after collision. In our hybrid QMC-iterative method, the statistical weight of each particle is now a vector of weights, similar to Equation \ref{eq:vecs}:
\begin{equation}
    \vec{w} = (w_1,w_2,\dots,w_G)^{T}.
\end{equation}
Therefore, we only need to multiply this vector by the scattering cross section (Eq.~\ref{eq:MatricesS}) to determine the scattering distribution.


\section{Implementation Details}
\label{sec:algorithm}

The algorithm was written in Julia, a scientific computing language that combines the compiler capabilities of C++ and the syntax of Matlab and Python. The code and primary documentation are available here \href{https://github.com/spasmann/iQMC.jl}{iQMC.jl} \cite{ctk:krylovqmc}. The Krylov linear solvers come from the Julia package \href{https://github.com/ctkelley/SIAMFANLEquations.jl}{SIAMFANLEQ.jl} \cite{ctk:siamfanl}. The documentation for these codes is in the \href{https://github.com/ctkelley/NotebookSIAMFANL}{Julia notebooks} \cite{ctk:notebooknl} and the book \cite{ctk:fajulia} that accompany the package. 

The results in Section \ref{sec:results} were generated using the Sobol Sequence as the LDS in \textit{QMC Sweep}. The Sobol Sequence generates nets with $2^m$ points and loses some of its balance properties if generated with a sample size that is not a power of 2 \cite{owen2022dropping}. Consequently, the number of particles $N$, varied by powers of 2 in all experiments. A brief convergence comparison between the Sobol, Halton, and Golden sequences can be seen in Problem \ref{sec:reeds}, Figure \ref{fig:lds_comp}. Finally, for an equivalent comparison to MC, the LDS was replaced with a fixed-seed pseudo-random number generator. By fixed-seed, we refer to a calculation where the random number seed is reset at the beginning of each iteration: this assures that the stochasticity of the MC algorithm is not an impediment to convergence.  An outline of the \textit{QMC Sweep} algorithm is described below (Algorithm \ref{alg:qmc_sweep}).

\vspace{0.5cm}

\begin{minipage}{0.66\textwidth}
\begin{algorithm}[H]
    \centering
    \caption{QMC Sweep ($\phi_\textrm{in}$)}\label{alg:qmc_sweep}
    \begin{algorithmic}[1]
		\STATE Initialize Low-Discrepancy-Sequence (LDS)
		\FOR {$i$ in $N$}
			\STATE Assign position and angle ($x_i$, $\mu_i$) based on the LDS
			\STATE Initialize weight 
			$w_i=\left(\Sigma_{s}\phi_{\textrm{in}}+q\right)N_xV/N$
			\FOR {$j$ in $N_x$}
				\STATE Move particle across $\textrm{Zone}_j$
				\STATE Tally($x, \mu, w$) (Eq: \ref{equation:tally})
				\STATE Update particle weight (Eq. \ref{equation:CWR})
			\ENDFOR
		\ENDFOR
		\STATE Return: $\phi_\textrm{out}$
    \end{algorithmic}
\end{algorithm}
\end{minipage}

\vspace{0.75cm}

As previously mentioned in Section \ref{sec:methods}, the described QMC-iterative algorithm does not require the explicit formation of the matrix $A$ for the linear solver. Instead, we compute the matrix vector product $A\phi$, as described in Algorithm \ref{alg:mvp}, from which the Krylov solvers can iterate.

\vspace{0.5cm}

\begin{minipage}{0.66\textwidth}
\begin{algorithm}[H]
    \centering
    \caption{Matrix Vector Product ($\phi_\textrm{in}$)}\label{alg:mvp}
    \begin{algorithmic}[1]
		\STATE $\textrm{b}=\textrm{QMC Sweep}(\vec{0})$
		\STATE $\textrm{mxv}=\textrm{QMC Sweep}(\phi_\textrm{in})$
		\STATE $\textrm{axv}=\phi_\textrm{in}-\textrm{mxv}-\textrm{b}$
		\STATE Return: axv
    \end{algorithmic}
\end{algorithm}
\end{minipage}

\vspace{0.5cm}

To evaluate the effect of varying the number of spatial cells ($N_x$) within a QMC simulation, a post-process spatial averaging technique was developed. Given a reference solution for the scalar flux with $N_{x\textrm{Ref}}$ spatial cells, experiments were run for problems \ref{sec:mg} and \ref{sec:reeds} with $N_x=(N_{x\textrm{Ref}})^{2^n}$. The resulting vectors can be averaged $\left[\log(N_x/N_{x\textrm{Ref}})/\log(2)\right]$ times to reduce all vectors to the length of the reference solution as seen in Algorithm \ref{alg:reduceflux}. 

\vspace{0.5cm}

\begin{minipage}{0.66\textwidth}
\begin{algorithm}[H]
    \centering
    \caption{Flux Spatial Average ($\phi_\textrm{in}$, $N_{x\textrm{Ref}}$)}\label{alg:reduceflux}
    \begin{algorithmic}[1]
        \STATE $\phi_{\textrm{out}} = \phi_\textrm{in}$
		\STATE $N_x=\textrm{length}(\phi_\textrm{out})$
		\STATE $I=\log(N_x/N_{x\textrm{Ref}})/\log(2)$
		\FOR {$i=1$ to $I$}
            \STATE $\textrm{LeftCells}=\phi_\textrm{out}[1:2:N_x-1]$
            \STATE $\textrm{RightCells} = \phi_\textrm{out}[2:2:N_x]$
            \STATE $\phi_\textrm{out} = (\textrm{RightCells} + \textrm{LeftCells})*0.5$
            \STATE $N_x=\textrm{length}(\phi_\textrm{out})$
		\ENDFOR
		\STATE Assert$\left( \textrm{length}(\phi_\textrm{out})=\textrm{length}(N_{x\textrm{Ref}}) \right)$
		\STATE Return: $\phi_\textrm{out}$
    \end{algorithmic}
\end{algorithm}
\end{minipage}

\vspace{0.75cm}

\section{Computational Results}
\label{sec:results}



\subsection{Problem 1: Multigroup in an Infinite Medium}
\label{sec:mg}

The first problem features 12, 70, and 618 group cross section data of high-density polyethylene (HDPE) generated with FUDGE \cite{mattoon2012generalized} in an infinite medium.  For a constant volumetric source $Q$, the analytic solution for the scalar flux is given by:
\begin{equation}
    \label{eq:mg_sol}
    \phi^\textrm{Sol} = (\Sigma_t - \Sigma_s)^{-1}\cdot Q.
\end{equation}

\begin{figure}[H]
\centerline{
\includegraphics[width=0.66\textwidth]{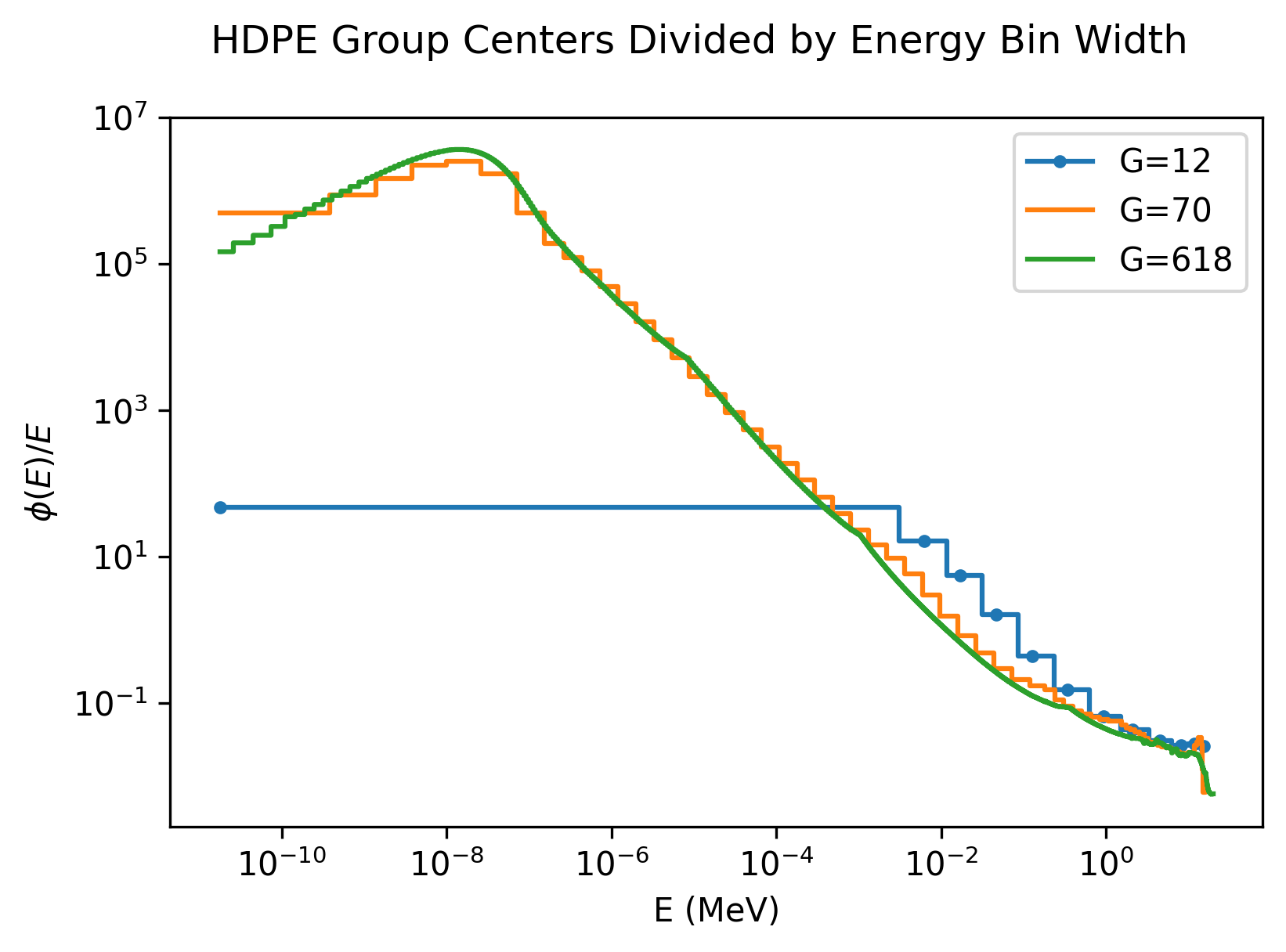}
}
\caption{\label{fig:mg_sol} Energy group centers $0.5(E_\textrm{Max} - E_\textrm{Min})$ (MeV) of total cross sections ($\Sigma_t$) from HDPE data, against the analytic solution for scalar flux (Equation \ref{eq:mg_sol}) divided by the energy bin width for 12, 17, and 618 group data sets.}
\end{figure}

To simulate an infinite medium, we placed an isotropic boundary source on each slab edge, where the source strength was held at the expected analytic solution. Figure~\ref{fig:mg_sol} shows the center of each energy group for the total cross section against the analytic solution for scalar flux from Eq.~\eqref{eq:mg_sol} divided by the energy bin width. 

\begin{figure}[H]
     \centering
     \begin{subfigure}[b]{0.48\textwidth}
         \centering
         \includegraphics[width=\textwidth]{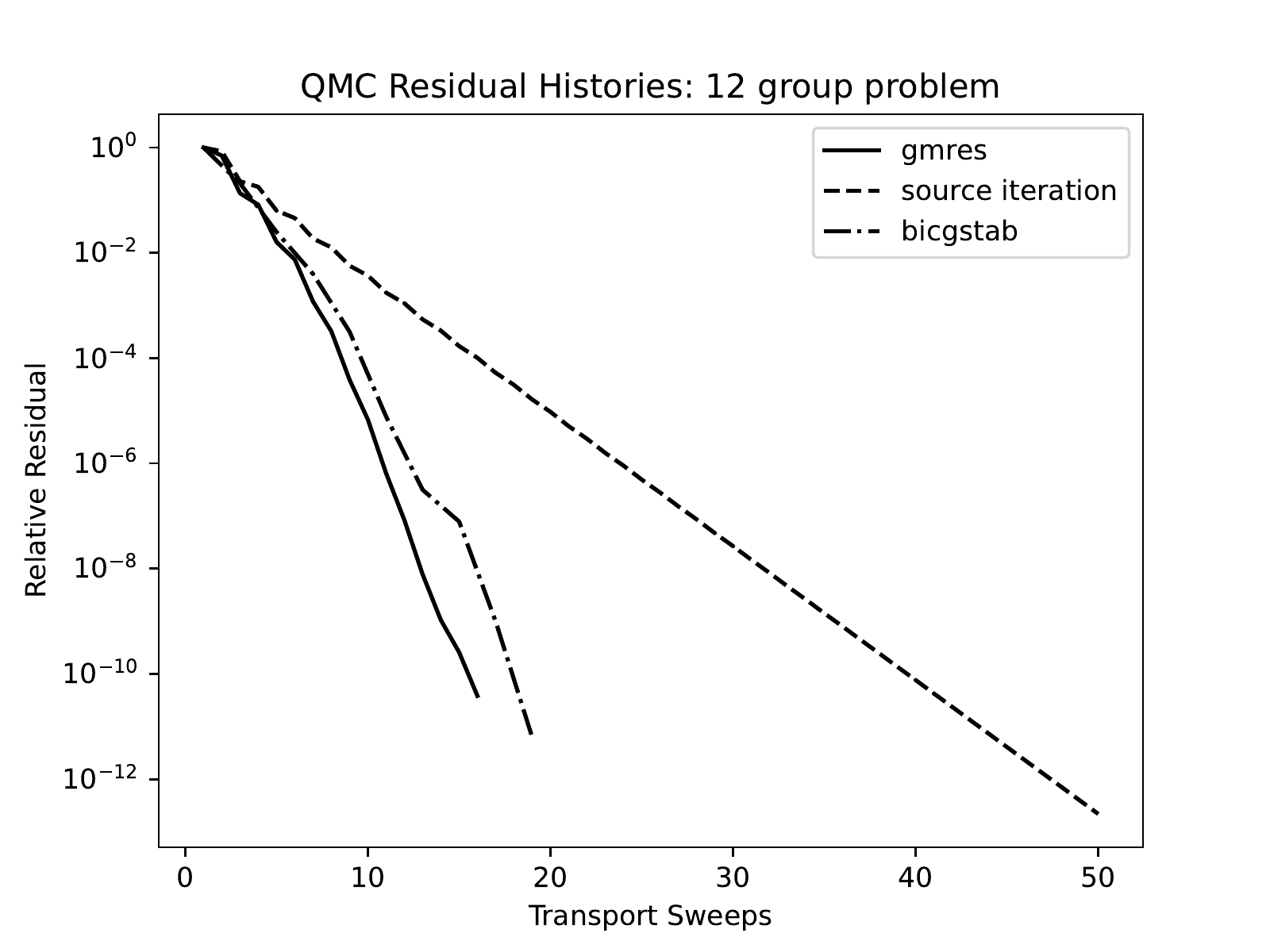}
         \caption{\label{fig:618group}}
     \end{subfigure}
     \hfill
     \begin{subfigure}[b]{0.48\textwidth}
         \centering
         \includegraphics[width=\textwidth]{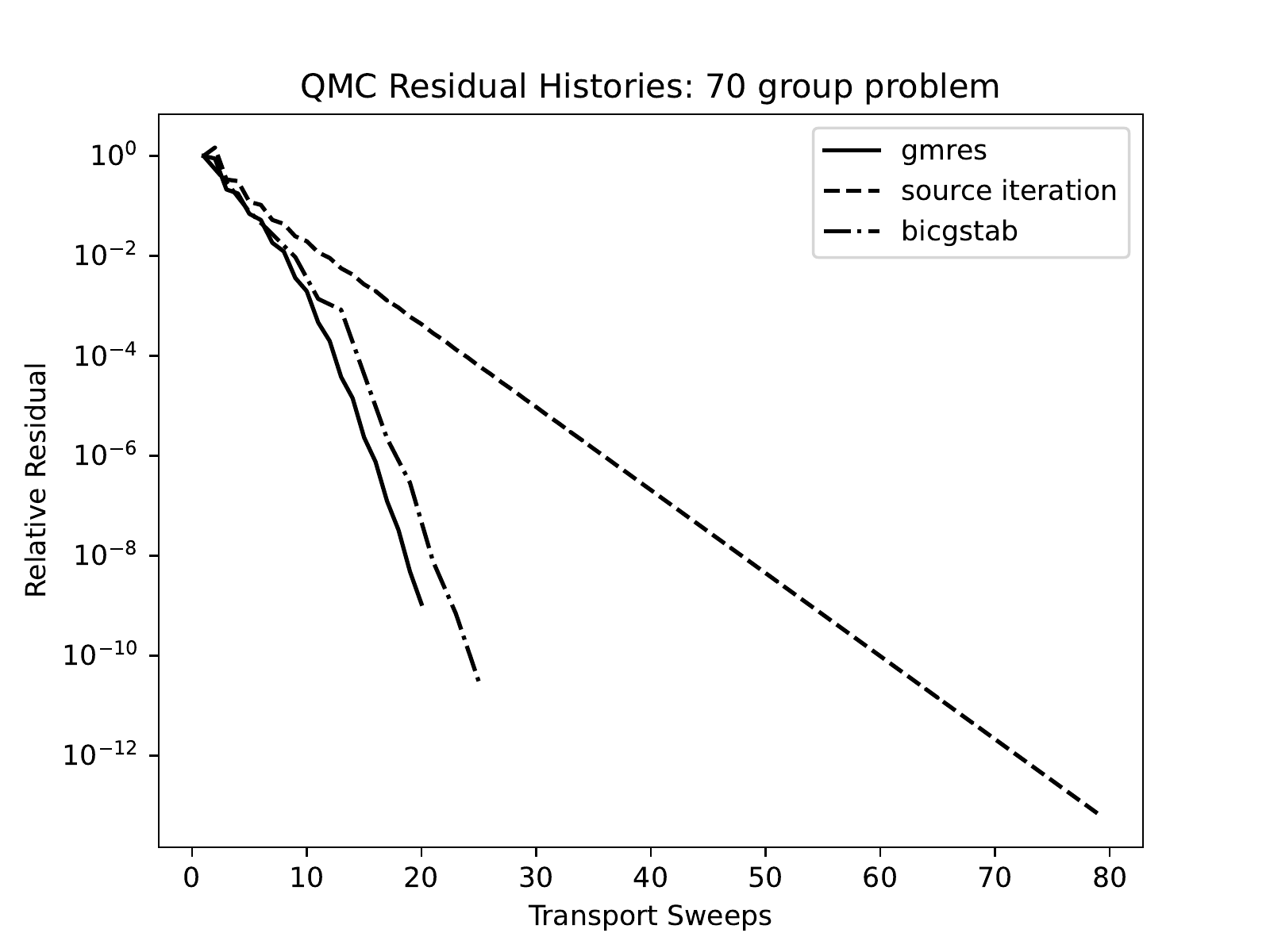}
         \caption{\label{fig:70group}}
     \end{subfigure}
     \\
     \begin{subfigure}[b]{0.48\textwidth}
         \centering
         \includegraphics[width=\textwidth]{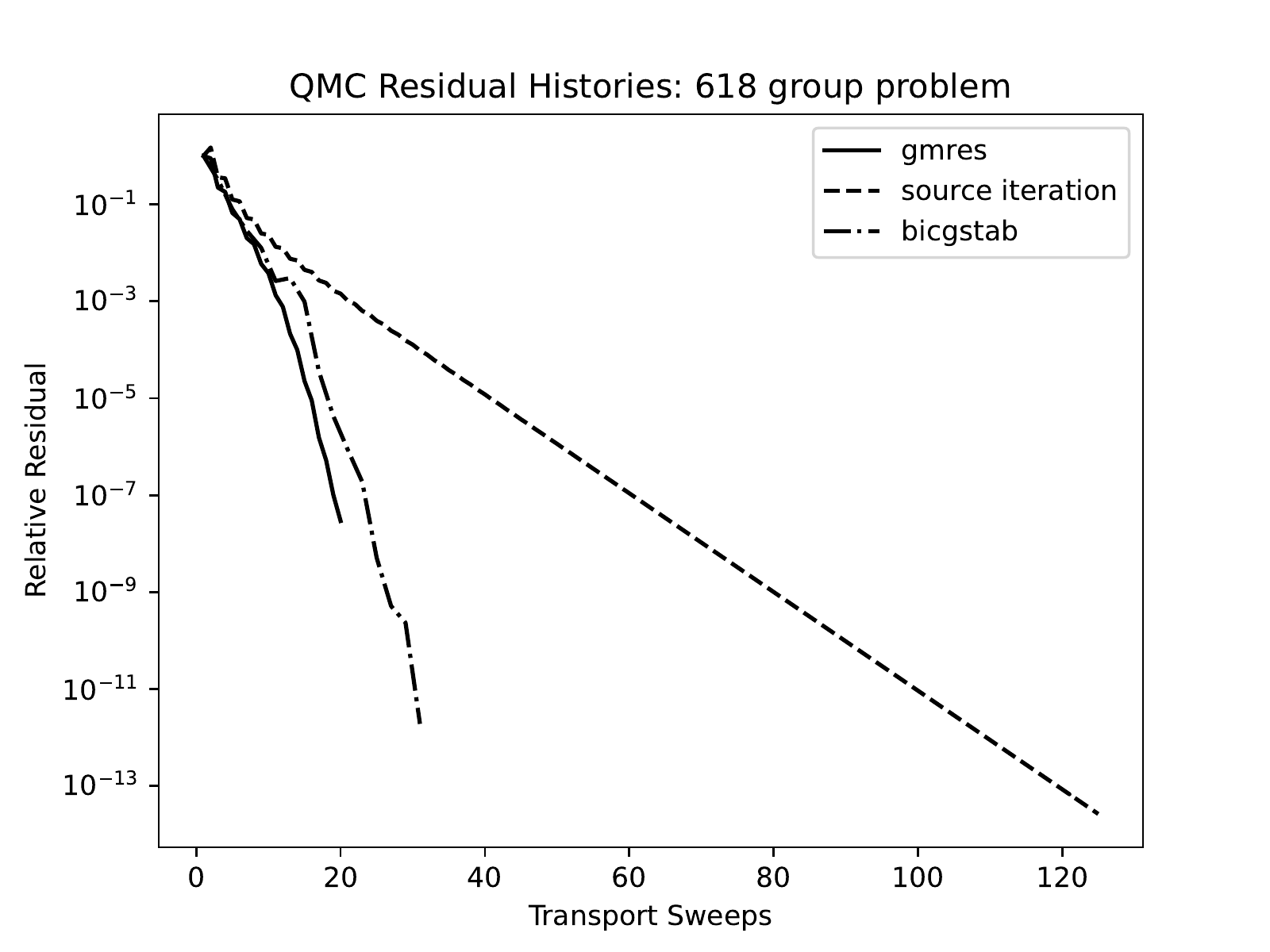}
         \caption{\label{fig:12group}}
     \end{subfigure}
        \caption{Multigroup relative residuals defined as $||\phi_n - \phi_{n-1}||$ for 12 (a), 70 (b), and 618 (c) group problems with $N=2048$.}
        \label{fig:mg_residuals}
\end{figure}

\begin{figure}[H]
\centerline{
\includegraphics[width=\textwidth]{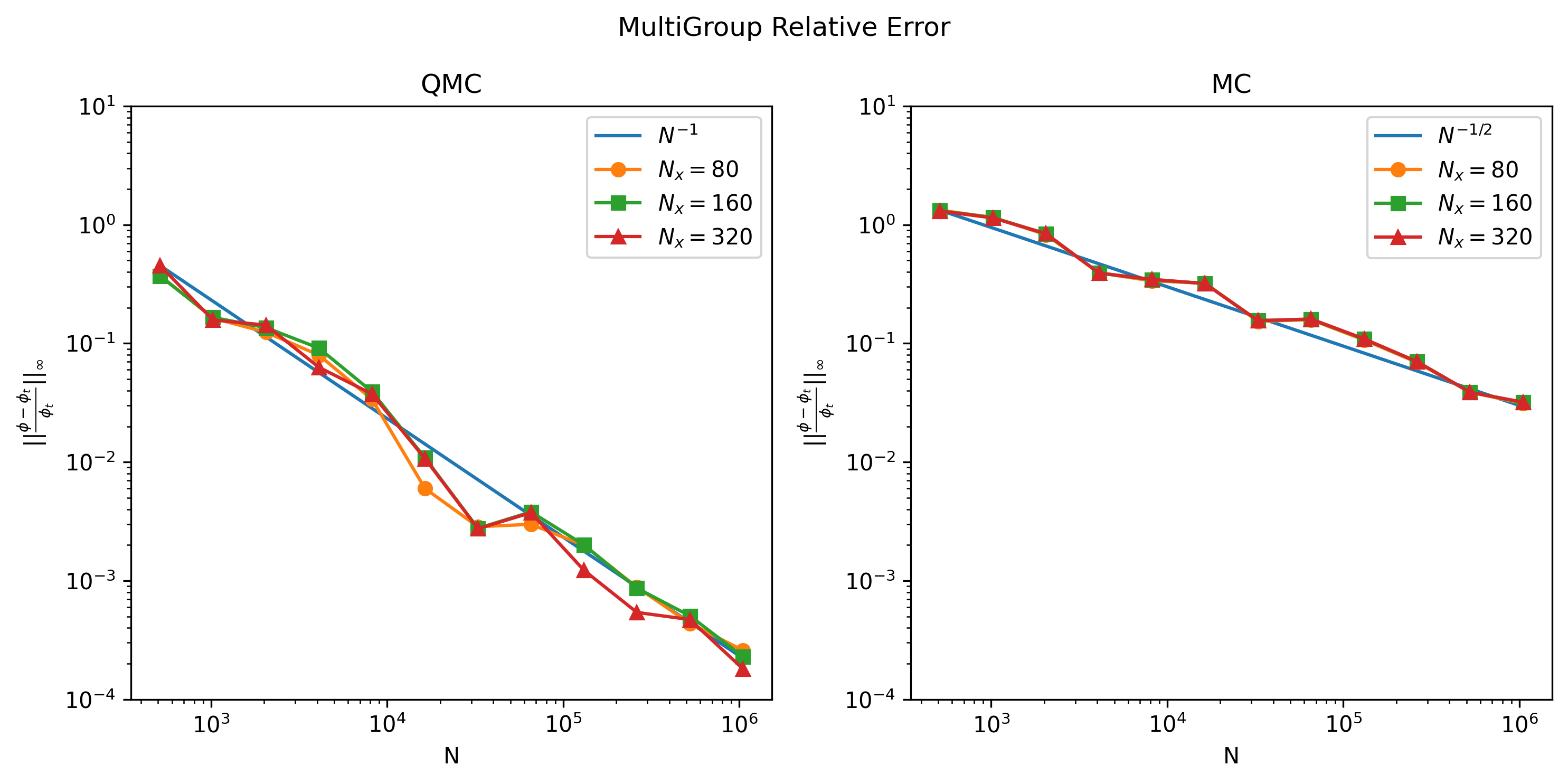}
}
\caption{\label{fig:mg_RE} Relative error from the 12 group QMC and MC scalar flux results compared to analytic solution for varying number of particles $N$ and spatial cells $N_x$. The QMC result achieve the expected $O(N^{-1})$ convergence, while the MC results converge at the standard $O(N^{-1/2})$.}
\end{figure}

From the convergence of relative residuals in Figure \ref{fig:mg_residuals}, it is observed that the Krylov methods require far fewer transport sweeps to achieve the same levels of convergence as the Source Iteration regardless of the number of groups. Figure \ref{fig:mg_RE} plots the residual, 
\begin{equation}\label{eq:mg_RE}
    R = 
\frac{ || \phi^\textrm{Sol} - \phi^{QMC} ||_\infty }{\phi^\textrm{Sol}},
\end{equation} 
against the number of particle histories $N$ for a given number of spatial cells $N_x$ along with the theoretical convergence goal of $O(N^{-1})$ for QMC and $O(N^{-1/2})$ for MC. 

Simulations were run with 80, 160, and 320 spatial cells and afterward results were run through the spatially averaged scalar flux post-processing technique from Algorithm \ref{alg:reduceflux} . This allows for comparison of all results to a solution $\phi^\textrm{Sol}$ with $N_x=80$. Because the solution is spatially constant per energy group, the results converge at the approximate rate of $O(N^{-1})$ for QMC and $O(N^{-1/2})$ for MC,  regardless of the number of spatial cells (Figure \ref{fig:mg_RE}).

\subsection{Problem 2: Reed's Multi-Media Problem}
\label{sec:reeds}

\begin{figure}[h]
\centerline{
\includegraphics[width=0.9\textwidth]{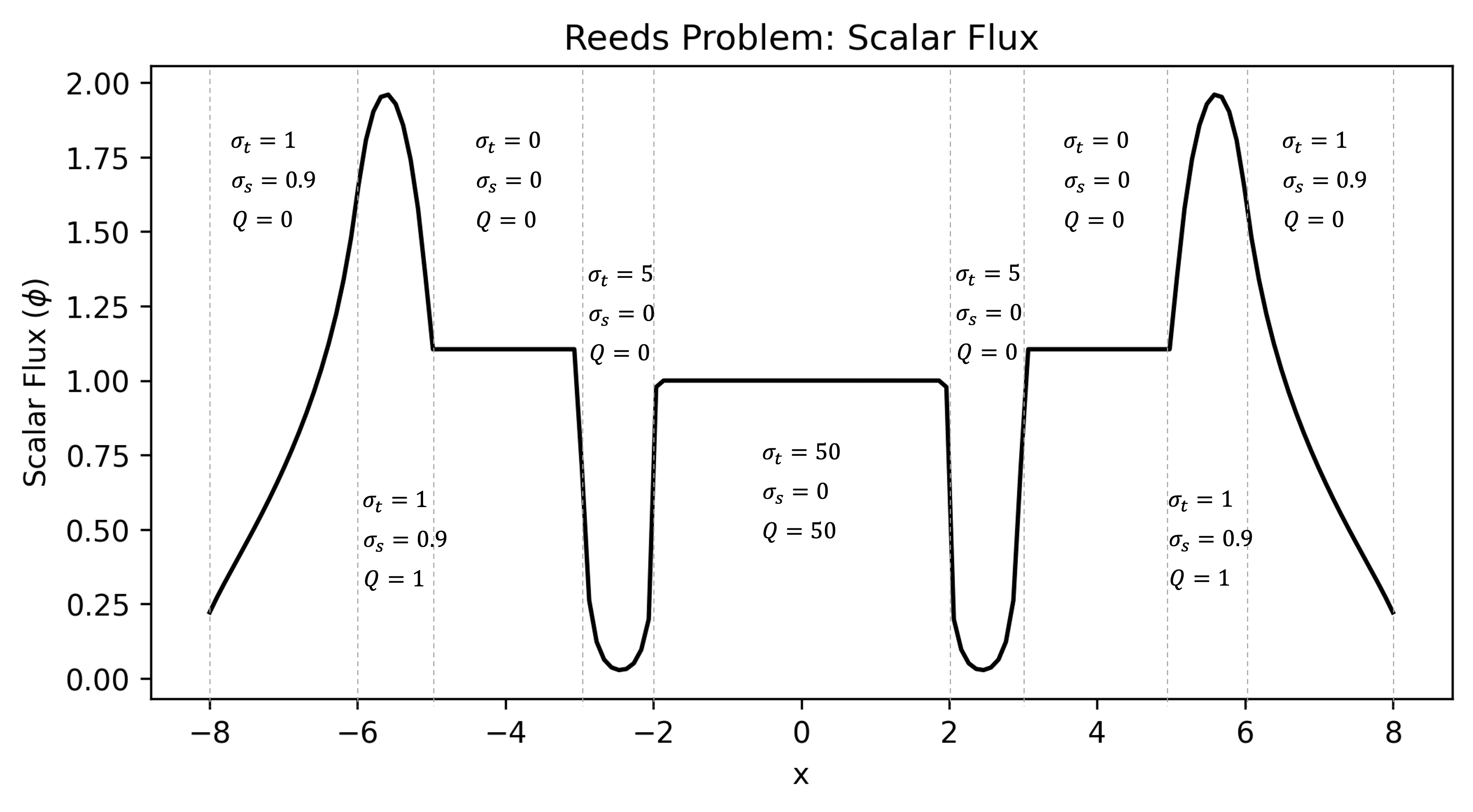}
}
\caption{\label{fig:reeds} MCDC solution for scalar flux $\phi$ with regional cross section data and source strength.}\end{figure}

The second problem, known as \textit{Reed's Problem}, is a mono-energetic multi-media problem in slab geometry \cite{Warsa2002}. The problem features 5 unique media that are reflected across the problem for a total of 9 regions, see Figure~\ref{fig:reeds}. To ensure that each spatial cell contained only one media, Reed's Problem was run so that $N_x$ was evenly divisible by 16, the range of the problem.

Reed's Problem was benchmarked using results from a $N=10^{10}$ analog Monte Carlo simulation from CEMeNT's Monte Carlo Dynamic Code (MCDC) \cite{Variansyah2022}. Additionally, in analyzing the relative error, it was observed that the solution to the problem approaches zero in multiple locations and this was drastically increasing the relative error as reported in the previous problem. Instead, for Reed's Problem we report the $L_\infty$ norm of the error as seen in Equation \ref{eq:reeds_err}. 

\begin{equation}\label{eq:reeds_err}
    R = || \phi^\textrm{Sol} - \phi^{QMC} ||_\infty.
\end{equation}

\begin{figure}[H]
\centerline{
\includegraphics[width=0.6\textwidth]{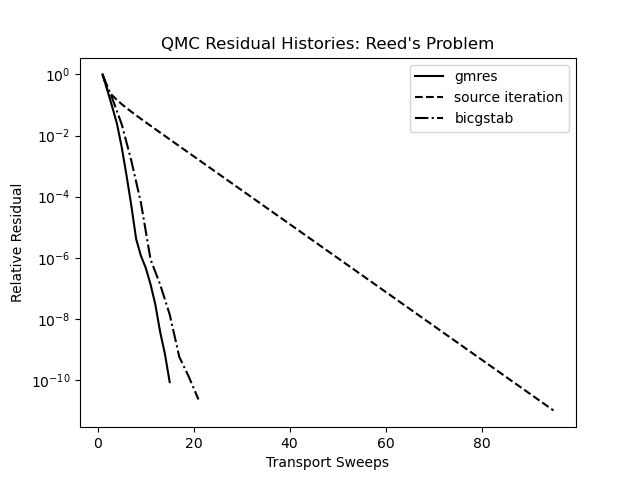}
}
\caption{\label{fig:reeds_residuals} Scalar flux relative residuals from Reed's Problem given $N=2048$ and $N_x=128$.}
\end{figure}

\begin{figure}[H]
\centerline{
\includegraphics[width=\textwidth]{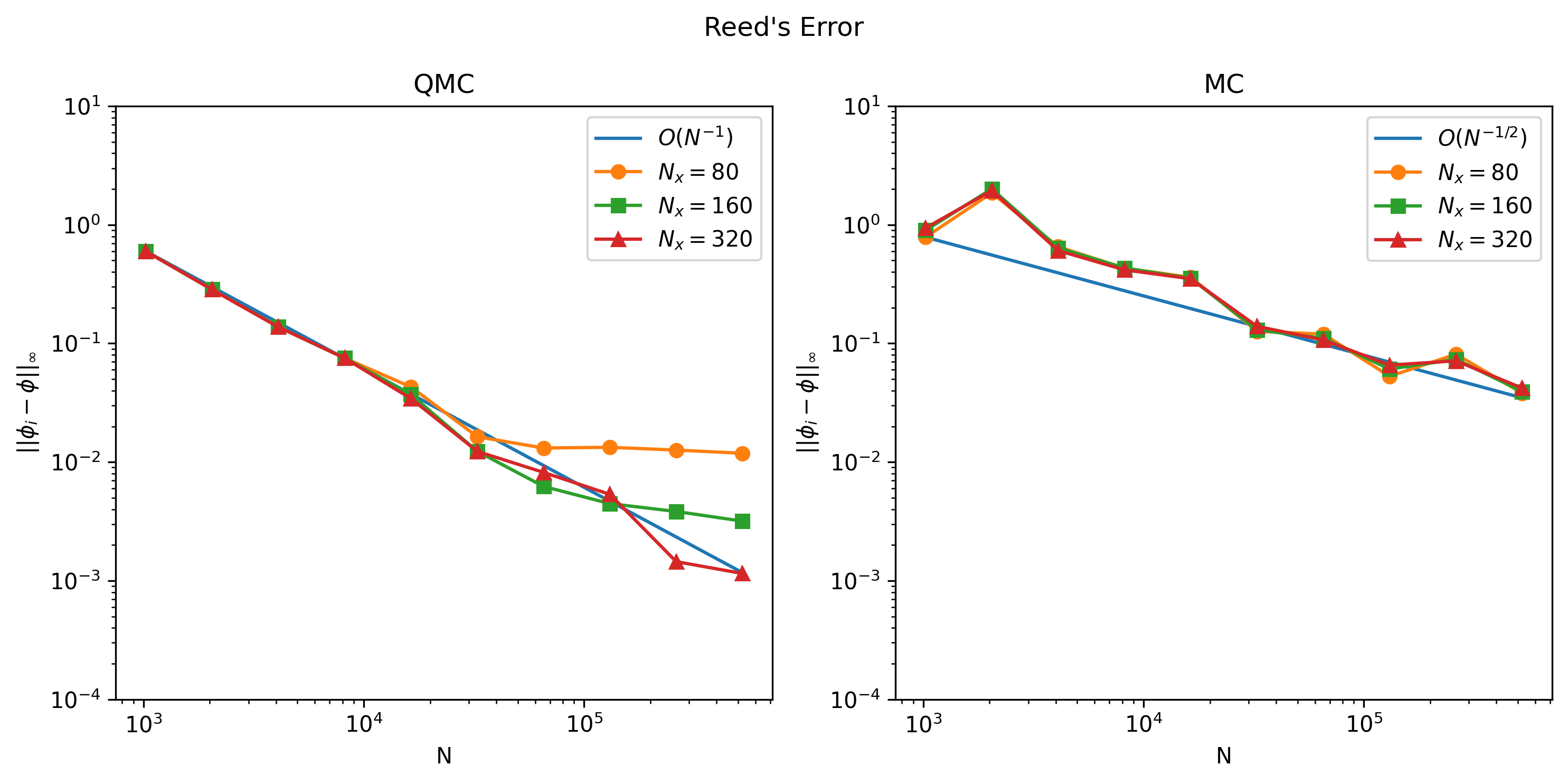}
}
\caption{\label{fig:reeds_re} $L_\infty$ of the absolute error of QMC scalar flux results compared to MCDC for varying number of particles $N$ and spatial cells $N_x$. The QMC results converge at $O(N^{-1})$ until limited by the spatial error. Increasing $N_x$ lowers the spatial error limit.}
\end{figure}

\begin{figure}[H]
\centerline{
\includegraphics[width=0.55\textwidth]{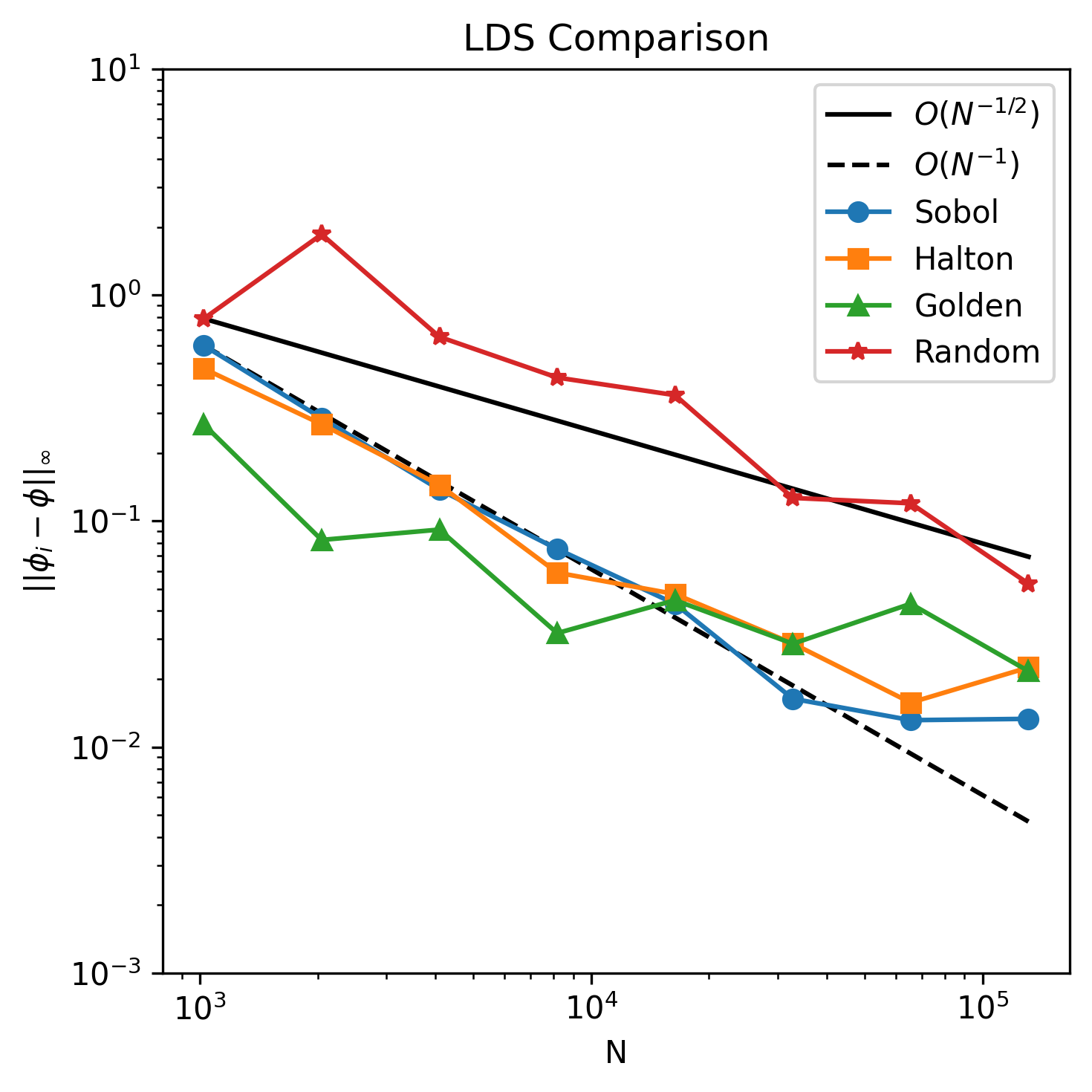}
}
\caption{\label{fig:lds_comp} Comparison of the Sobol, Halton, and Golden Sequences along with results from a pseudo-random number generator for $N_x=80$. The Golden Sequence achieves the best accuracy for low particle histories but is eventually out performed by the Sobol Sequence.}
\end{figure}

Again, it is observed that the Krylov solvers far outperform Source Iteration, requiring no less than one-quarter the number of iterations to converge (Figure~\ref{fig:reeds_residuals}). Unlike problem \ref{sec:mg} however, the effects of increasing the number of spatial cells and utilizing the spatially averaged scalar flux algorithm are clearly seen (Figure \ref{fig:reeds_re}). Our \textit{QMC Sweep} currently utilizes a flat source for each cell, and beyond a certain number of particle histories $N$, the convergence is limited by spatial error determined by $N_x$. As the number of spatial cells is increased, the spatial error is reduced and the QMC can continue to converge at the $O(N^{-1})$ rate. Note, this effect is not observed in the MC results, because they did not reach the spatial error limit for $N_x=80$, near $10^{-2}$. Finally, Figure \ref{fig:lds_comp} shows the results from a simulation with $N_x=80$ using the Sobol, Halton, and Golden sequences and a pseudo-random number generator (MC). The MC results perform as expected and the three LDS perform rather similarly. The Golden sequence achieves a lower error for the first few particle counts but the Sobol Sequence results plateau at a lower error and ultimately achieves the greatest accuracy.

\subsection{Problem 3: Isotropic Boundary Source}
\label{sec:garcia}

The third and final computational experiment solves the mono-energetic problem from Garcia et al. \cite{cesinh}, outlined in Table~\ref{tab:garcia}. Here, the scattering cross section is spatially dependent, defined by $\Sigma_s = e^{-x/s}$, and we considered two cases: $s=1$ and $s=\infty$. Note that $s=\infty$ is equivalent to a constant scattering cross section and therefore is the \textit{harder} of the two scenarios as it involves more scattering and therefore more iterations for the scattering source to converge.

\begin{figure}[H]
\centerline{
\includegraphics[width=3.5in]{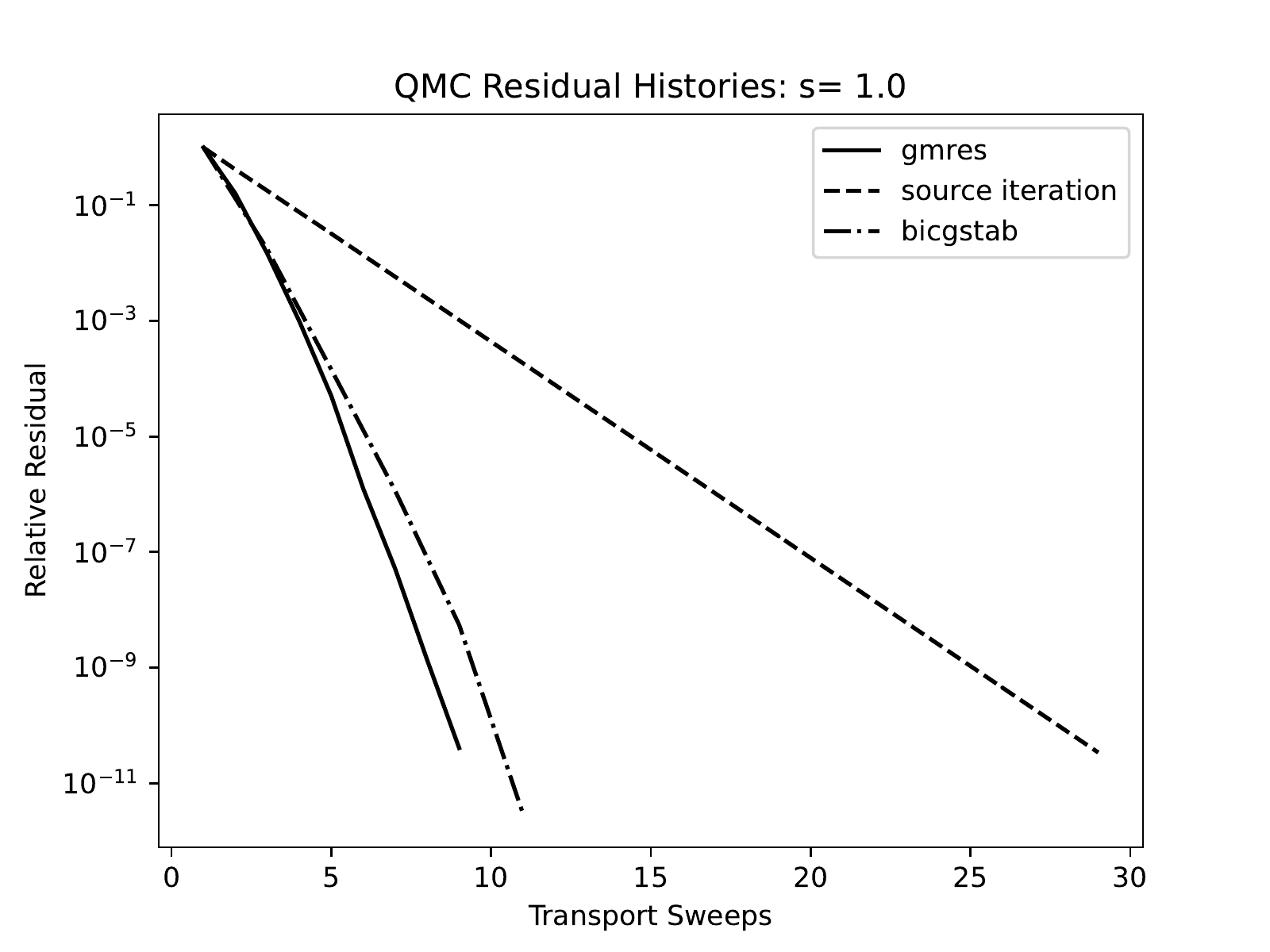}
}
\caption{\label{fig:easy} Scalar flux relative residuals for $s=1$ given parameters from Table ~\ref{tab:garcia} and $N=2048$ and $N_x = 100$.}
\end{figure}

\begin{figure}[H]
\centerline{
\includegraphics[width=3.5in]{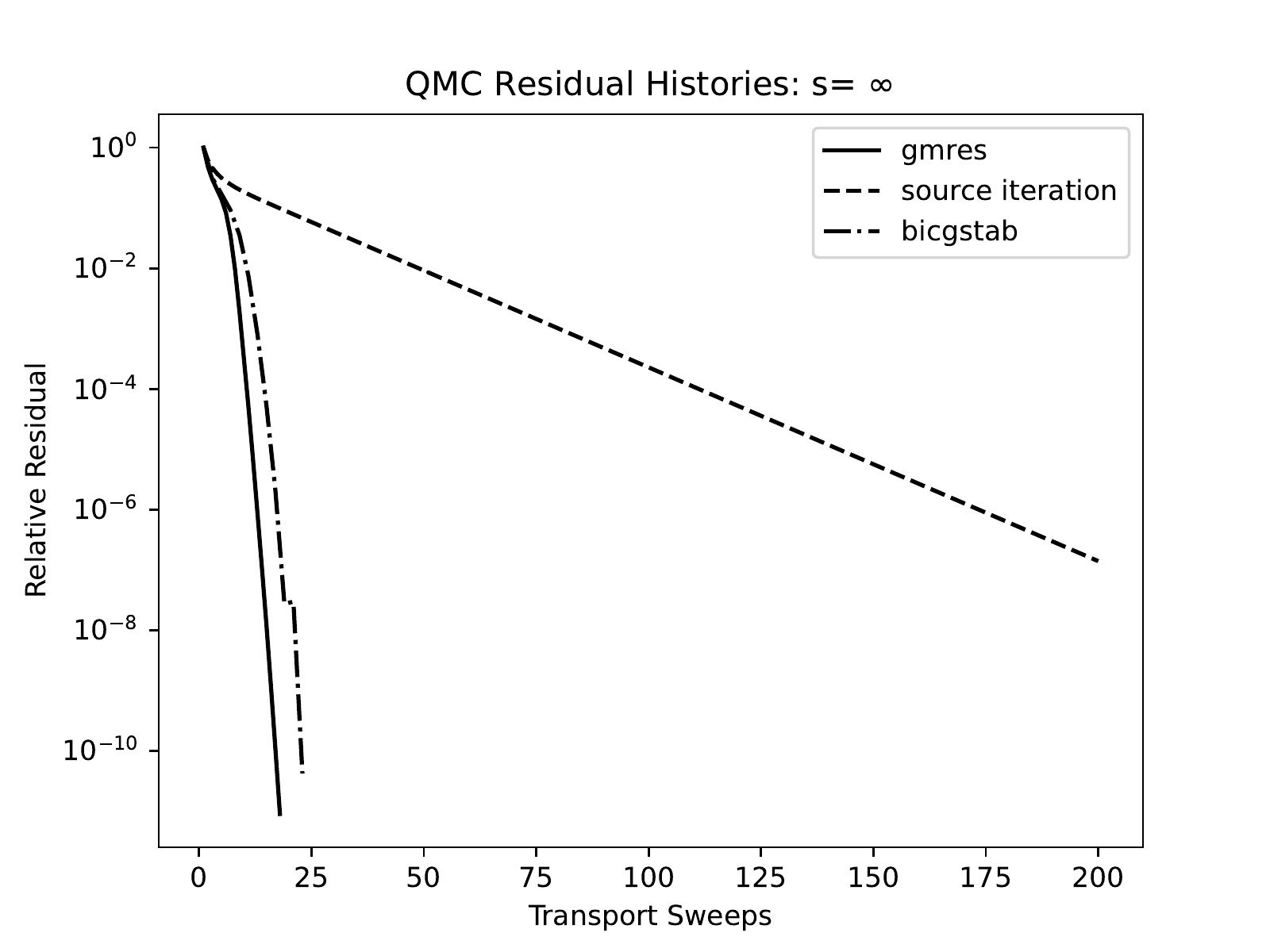}
}
\caption{\label{fig:hard} Scalar flux relative residuals for $s=\infty$ given parameters from Table ~\ref{tab:garcia} and $N=2048$ and $N_x=100$.}
\end{figure}

First, we solve the QMC linear problem with $N=2048$ particles and $N_x=100$ spatial cells. Similar to Problems \ref{sec:reeds} and \ref{sec:mg}, Figure~\ref{fig:easy} shows that for an exponentially decaying scattering cross section ($s=1$) the Krylov iterations take fewer than a third of the number of transport sweeps than that of the SI for a relative residual of $10^{-9}$. While Figure~\ref{fig:hard} shows that for the constant scattering cross section ($s=\infty$) the Krylov iterations took less than 25 iterations to reach a relative error of $10^{-6}$ while the SI required nearly 200 iterations. 

\begin{table}[h]
\centering
\caption{Parameters for fixed boundary source, slab geometry, simulation from Garcia et al. \cite{cesinh}}
\label{tab:garcia}
\centerline{
\begin{tabular}{c | c}
\hline
 Parameter & Value \\ 
\hline
 $\Sigma_t$ & 1 \\  
 $\Sigma_s(x)$ & $e^{-x/s}$ \\
 $\tau$ & $5$ \\
 $\psi_l(\mu)$ & 1 \\
 $ \psi_r(\mu)$ & 0 \\
 $N_x$ & 50 \\
 $q(x)$ & 0 \\
\hline
\end{tabular}
}
\end{table}


\subsubsection{Validation and calibration study}
\label{validation-and-calibration-study}

\begin{figure}[H]
\centerline{
\includegraphics[width=\textwidth]{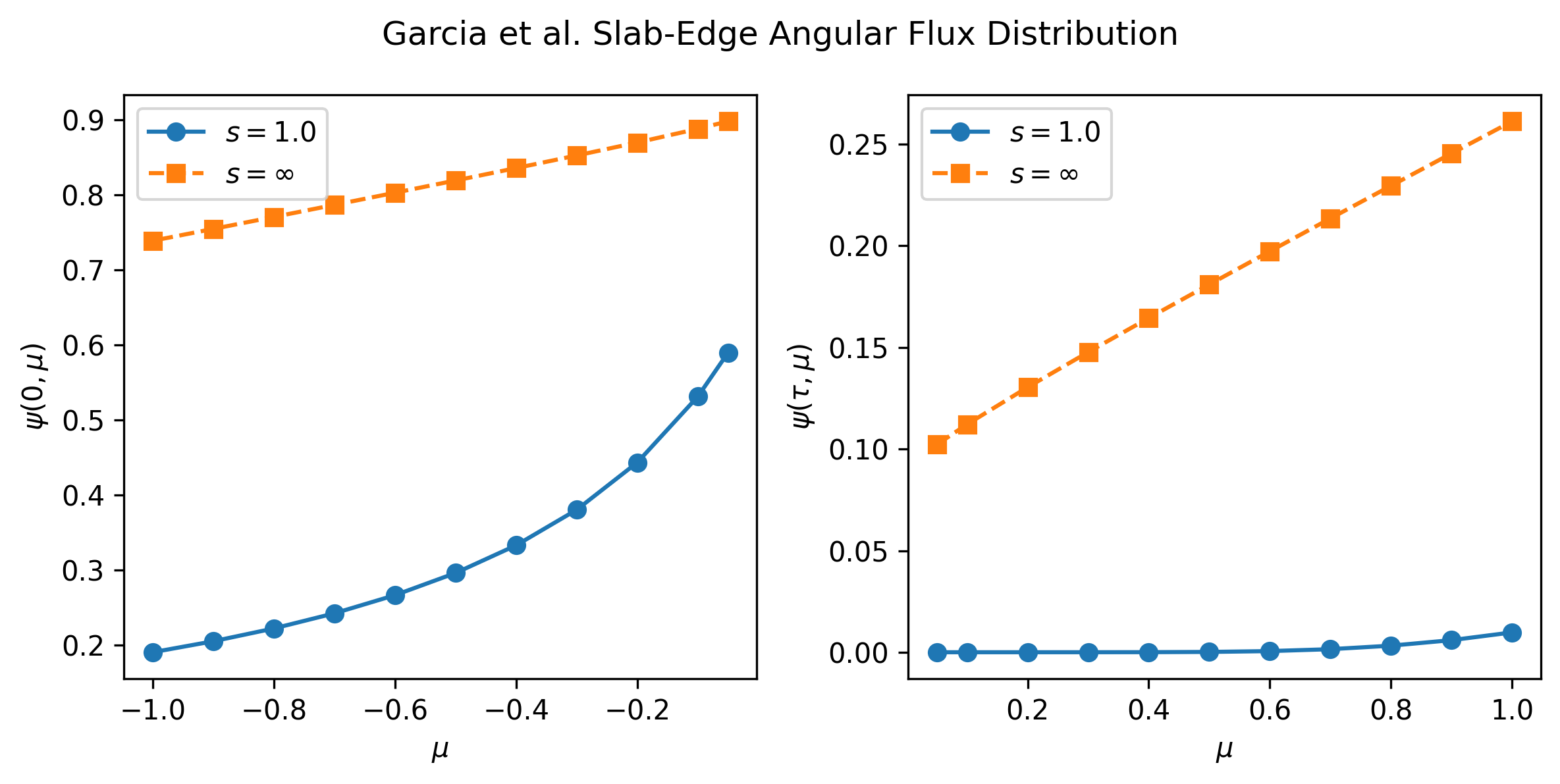}
}
\caption{\label{fig:garcia_sol} Angular flux exit distributions provided by Garcia et al. \cite{cesinh}. Recreated using scalar flux data from QMC and subsequent angular flux exit distributions from an $S_N$ approximation }
\end{figure}

We conclude this problem with a validation study given results from \cite{cesinh} that are angular flux exit distributions accurate to six figures, see Figure~\ref{fig:garcia_sol}. We duplicated the results by obtaining the cell-average scalar flux from the QMC simulation, for $N=2048$ and $N_x=100$. Then used a single $S_N$ transport sweep to recover the exit distributions. We report the corresponding results from \cite{cesinh} in Tables~\ref{tab:cesone} and \ref{tab:cesinf}. The exit distributions, as is clear from Table~\ref{tab:cesone}, can vary by five orders of magnitude. Even so, the results from QMC agree with the benchmarks to roughly two figures.

\begin{table}[h]
\centering
\caption{Angular flux exit Distributions from Garcia et al. and the $S_N$ QMC sweep (with $N=2048$ and $N_x=100$) for $s=1$.}
\label{tab:cesone}
\begin{tabular}{lllll}
 & \multicolumn{2}{c}{Garcia/Siewert}
 & \multicolumn{2}{c}{QMC}\\
\hline
$\mu$ &$\psi(0, -\mu)$ &$\psi(\tau, \mu)$ &$\psi(0, -\mu)$ &$\psi(\tau, \mu)$ \\
\hline
 0.05 &    5.89664e-01 &    6.07488e-06 &    6.07035e-01 &    5.91908e-06   \\
 0.10 &    5.31120e-01 &    6.92516e-06 &    5.47466e-01 &    6.74075e-06   \\
 0.20 &    4.43280e-01 &    9.64232e-06 &    4.57064e-01 &    9.35453e-06   \\
 0.30 &    3.80306e-01 &    1.62339e-05 &    3.92223e-01 &    1.56108e-05   \\
 0.40 &    3.32964e-01 &    4.38580e-05 &    3.43481e-01 &    4.13721e-05   \\
 0.50 &    2.96090e-01 &    1.69372e-04 &    3.05510e-01 &    1.58622e-04   \\ 
 0.60 &    2.66563e-01 &    5.73465e-04 &    2.75098e-01 &    5.39514e-04   \\ 
 0.70 &    2.42390e-01 &    1.51282e-03 &    2.50192e-01 &    1.43257e-03   \\ 
 0.80 &    2.22235e-01 &    3.24369e-03 &    2.29422e-01 &    3.08975e-03   \\ 
 0.90 &    2.05174e-01 &    5.96036e-03 &    2.11837e-01 &    5.70555e-03   \\ 
 1.00 &    1.90546e-01 &    9.77123e-03 &    1.96756e-01 &    9.39189e-03   \\ 
\hline
\end{tabular}
\end{table}

\begin{table}[h]
\centering
\caption{Angular flux exit Distributions from Garcia et al. and the $S_N$ QMC sweep (with $N=2048$ and $N_x=100$) for $s=\infty$.}
\label{tab:cesinf}
\begin{tabular}{lllll}
 & \multicolumn{2}{c}{Garcia/Siewert}
 & \multicolumn{2}{c}{QMC}\\
\hline
$\mu$ &$\psi(0, -\mu)$ &$\psi(\tau, \mu)$ &$\psi(0, -\mu)$ &$\psi(\tau, \mu)$ \\
\hline
 0.05 &    8.97798e-01 &    1.02202e-01 &    9.06050e-01 &    1.03680e-01   \\ 
 0.10 &    8.87836e-01 &    1.12164e-01 &    8.95849e-01 &    1.13695e-01   \\ 
 0.20 &    8.69581e-01 &    1.30419e-01 &    8.76487e-01 &    1.31907e-01   \\ 
 0.30 &    8.52299e-01 &    1.47701e-01 &    8.58937e-01 &    1.49245e-01   \\ 
 0.40 &    8.35503e-01 &    1.64497e-01 &    8.42195e-01 &    1.66128e-01   \\ 
 0.50 &    8.18996e-01 &    1.81004e-01 &    8.25870e-01 &    1.82734e-01   \\ 
 0.60 &    8.02676e-01 &    1.97324e-01 &    8.09780e-01 &    1.99151e-01   \\ 
 0.70 &    7.86493e-01 &    2.13507e-01 &    7.93834e-01 &    2.15421e-01   \\ 
 0.80 &    7.70429e-01 &    2.29571e-01 &    7.77997e-01 &    2.31558e-01   \\ 
 0.90 &    7.54496e-01 &    2.45504e-01 &    7.62269e-01 &    2.47547e-01   \\ 
 1.00 &    7.38721e-01 &    2.61279e-01 &    7.46673e-01 &    2.63362e-01   \\ 
\hline
\end{tabular}
\end{table}

\begin{figure}[H]
\centerline{
\includegraphics[width=\textwidth]{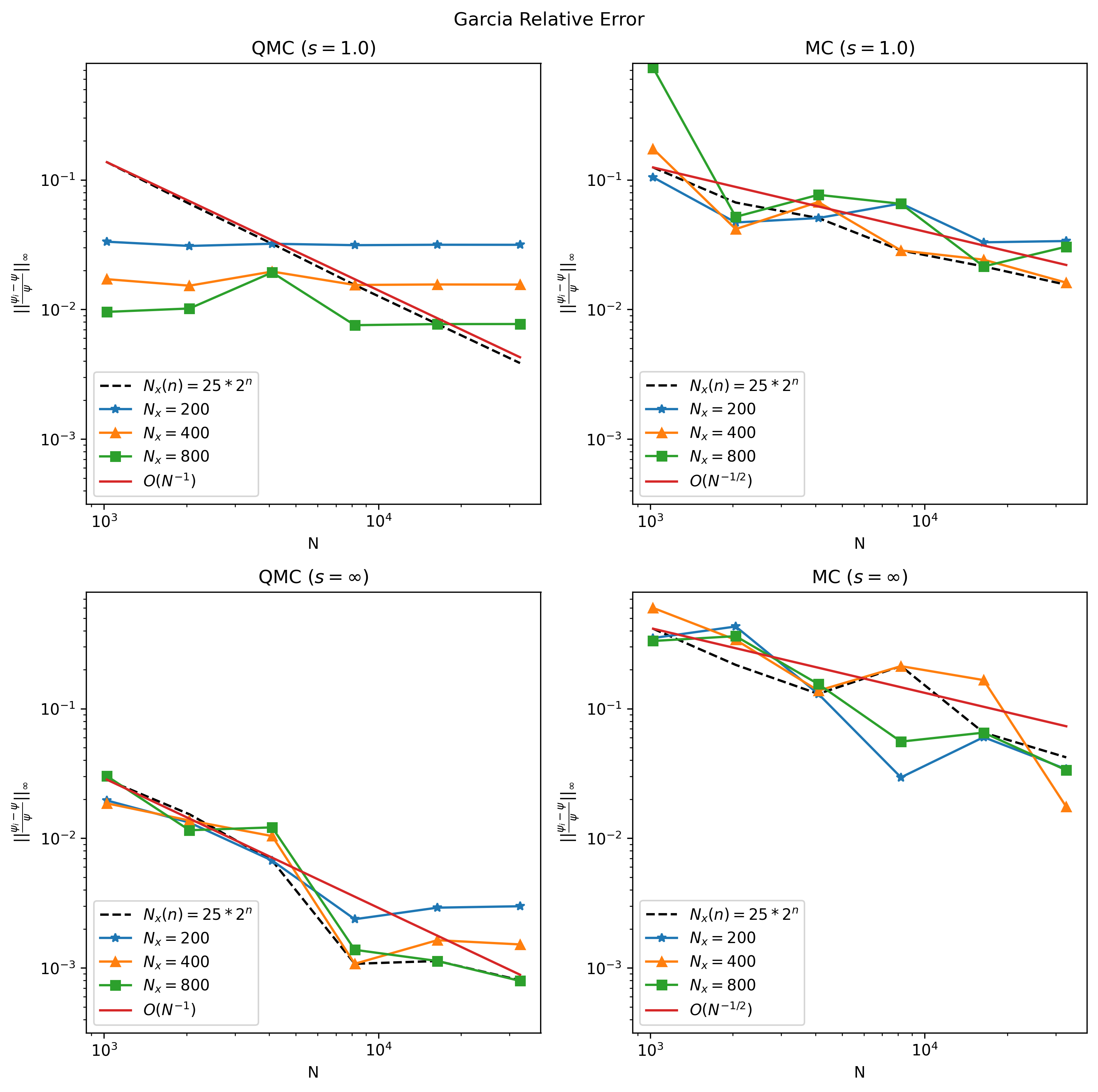}}
\caption{\label{fig:garcia_RE} Relative error of fixed-seed QMC and MC results for $s=1$ and $s=\infty$ compared to $S_N$ solution for varying number of particles $N$ and spatial cells $N_x$. Increasing the spatial cells and particle count simultaneously in QMC simulations achieves $O(N^{-1})$ convergence.}
\end{figure}

In Tables \ref{tab:bigtab1} and \ref{tab:bigtabinf} we look at the relative errors $R$ (Equations \ref{eq:R}, \ref{eq:RL}, \ref{eq:RR}) in the QMC exit distributions as compared to a highly accurate $S_N$ result. We compensate for the widely varying scales by tabulating, for each value of $N$ and $N_x$. Finally, similar to the previous two problems we plot the residual $R$ for varying $N$ and $N_x$ in Figure \ref{fig:garcia_RE}. However, unlike the previous two problems the scalar flux averaging algorithm was \textit{not} applied. In the QMC case of $s=1.0$ we notice that if the number of spatial cells are increased \textit{with} the number of particles, the $O(N^{-1})$ convergence is achieved. For $s=\infty$ the QMC results converge irrespective of the number of spatial cells. 

\begin{equation} \label{eq:R}
    R = \max(R^0, R^\tau)
\end{equation}
where
\begin{equation}\label{eq:RL}
    R^0 = \max_\mu
\frac{ | \psi^{SN}(0,-\mu) - \psi^{QMC}(0,-\mu) | }{\psi^{SN}(0,-\mu) },
\end{equation}
and
\begin{equation}\label{eq:RR}
    R^\tau = \max_\mu
\frac{ | \psi^{SN}(\tau,\mu) - \psi^{QMC}(\tau,\mu) | }{\psi^{SN}(\tau,\mu) }.
\end{equation}

\begin{table}[h]
\centering
\caption{Exit Distribution Errors ($R$): $s = 1.0$, for varying the number of spatial cells ($N_x$) and particles per transport sweep ($N$).}
\label{tab:bigtab1}
\begin{tabular}{l|lllll} 
\hline
 $N_x$ \textbackslash N &     1024 &     2048 &     4096 &     8192 &    16384 \\ 
\hline 
50   & 1.36975e-01& 1.34260e-01& 1.35123e-01& 1.35328e-01& 1.35242e-01   \\ 
100  & 6.09631e-02& 6.35764e-02& 6.46191e-02& 6.48898e-02& 6.48536e-02   \\ 
200  & 3.77223e-02& 3.12496e-02& 3.12005e-02& 3.17337e-02& 3.16710e-02   \\ 
400  & 1.71316e-02& 1.52618e-02& 1.96221e-02& 7.56867e-03& 7.71669e-03   \\ 
800  & 9.58555e-03& 1.01782e-02& 1.93042e-02& 7.56867e-03& 7.71669e-03   \\ 
1600 & 7.18716e-03& 1.20870e-02& 2.19512e-02& 3.63519e-03& 3.85206e-03   \\ 
3200 & 5.11139e-03& 1.28920e-02& 2.19849e-02& 2.36739e-03& 1.82554e-03   \\ 
\hline 
\end{tabular} 
\end{table}

\begin{table}[h]
\centering
\caption{Exit Distribution Errors ($R$): $s = \infty$, for varying the number of spatial cells ($N_x$) and particles per transport sweep ($N$).}
\label{tab:bigtabinf}
\begin{tabular}{l|lllll} 
\hline
 $N_x$ \textbackslash N &     1024 &     2048 &     4096 &     8192 &    16384 \\ 
\hline 
50   & 2.83725e-02& 1.22374e-02& 1.19906e-02& 1.15398e-02& 1.15951e-02   \\ 
100  & 2.41784e-02& 1.53219e-02& 6.81242e-03& 5.67372e-03& 5.90660e-03   \\ 
200  & 1.95960e-02& 1.32236e-02& 6.73505e-03& 2.37596e-03& 2.91740e-03   \\ 
400  & 1.86114e-02& 1.37280e-02& 1.04038e-02& 1.07665e-03& 1.63724e-03   \\ 
800  & 3.02170e-02& 1.15453e-03& 1.21401e-02& 1.37999e-03& 1.13099e-03   \\ 
1600 & 2.26681e-02& 3.93529e-02& 1.78938e-02& 1.17129e-03& 1.17343e-03   \\ 
3200 & 3.30886e-02& 5.40682e-02& 2.52993e-02& 1.48919e-03& 1.51562e-03   \\ 
\hline 
\end{tabular} 
\end{table}

\pagebreak
\section{Conclusion}
\label{sec:conclusion}

We have described a iQMC, a general purpose iterative - Quasi-Monte Carlo method for solving the neutron transport equation. The use of iterative solvers and a continuous weight absorption technique provide a well suited application for QMC which provides an enhanced convergence rate of $O(N^{-1})$ compared to the $O(N^{-1/2})$ of standard Monte Carlo simulation. Additionally, the use of advanced iterative solvers like the Krylov methods GMRES and BiCGSTAB provide greatly increased convergence of residuals. The benefits of this algorithm were observed on all three 1-D test problems where the hybrid method provided both more accurate and efficient solutions. Future work will look to see if these benefits are maintained on more difficult and complex problems including critical eigenvalue, time-dependent, 2D, and 3D problems. The massive parallelism inherent to Monte Carlo combined with iQMC's particle tracing and vectorized multigroup methods suggest the it would benefit greatly from parallel implementation on advanced architectures and GPUs.

\section*{Acknowledgments}

This work was funded by the Center for Exascale Monte-Carlo Neutron Transport (CEMeNT) a PSAAP-III project funded by the Department of Energy, DE-NA003967, 
and supported by National Science Foundation Grants
DMS-1745654,
and
DMS-1906446.

\pagebreak
\bibliographystyle{ans_js}
\bibliography{bibliography}

\end{document}